\documentclass[showpacs,preprintnumbers,amsmath,amssymb,prb,twocolumn]{revtex4}

\begin{document}
\title{Magnetic monopoles in a charged two-condensate Bose-Einstein system}
\author{Ying Jiang}
\email{jiang@thp.uni-koeln.de}
\affiliation{Institut f\"ur Theoretische Physik, Universit\"at zu K\"oln,
Z\"ulpicherstr.77, 50937 K\"oln, Germany}
\begin{abstract}
We propose that a charged two-condensate Bose system possesses
point-like topological defects which can be interpreted as
magnetic monopoles. By making use of the $\phi$-mapping theory,
the topological charges of these magnetic monopoles can be
expressed in terms of the Hopf indices and Brouwer degree of the
$\phi$-mapping.
\end{abstract}

\maketitle

Superconductivity in magnesium diboride (MgB$_2$) with a
remarkably high $T_c = 39\; {\rm K}$ was recently reported by
Nagamatsu {\it et al.} \cite{nature} Since then, great attention
has been directed towards understanding the detailed nature of
superconductivity in this simple intermetallic compound. Because
the superconductivity of MgB$_2$ possesses an isotope effect
consistent with the phonon-mediated electron pairing of the BCS
theory, but with an extremely high critical temperature, it has
reopened the question of the maximum $T_c$ that can be produced by
that mechanism . \cite{wang1,bouquet2,4,55} Furthermore, it raises
a new question: what is the mechanism of the superconductivity of
MbB$_2$, and, in particular, whether this is a one- or two-gap
superconductor.

The evidence for multiple gap structure involves tunneling measurements
of the gap.
Values of $2\Delta/k_BT_c$ ranging from 1.2 to 4 have been reported
\cite{liu1}. The values below the BCS weak-coupling limit  of 3.5  have been
attributed to surface effects, but the best-quality spectra \cite{88} show
a very clean gap with the number equal to 1.25. Sharvin contact measurements
\cite{99} reveal a gap at 4.3 meV ($2\Delta/k_BT_c= 2.6$), and additional
structures at $2\Delta/k_BT_c=1.5$ and 3, raising
the possibility of multiple gaps. Careful analysis of the temperature
and magnetic-field dependence of the specific heat suggests multiple gap
structure as well \cite{wang1,bouquet2}.
In fact, there exist two gaps of different magnitude associated
with different bands in MgB$_2$. Their ratio is estimated as $r=\Delta_0 ^ S/
\Delta_0 ^ L \sim 0.3-0.4$ \cite{szabo1}
 where the larger gap $\Delta_0 ^ L$ is associated
with the two-dimensional $\sigma$-bands and the smaller gap $\Delta_0 ^ S$
with the three-dimensional $\pi$-bands \cite{tewordt1}.

Moreover, the experiment of scanning tunneling spectroscopy (STS)
measurements on single
crystal MgB$_2$ shows that the coherent length in the $\pi$ band is
approximately 50 nm \cite{eskildsen1}
 which is much
larger than an estimate which one would obtain from a standard GL formula.
All these spark renewed interest to two-gap superconductivity.

Two-gap
superconductivity is now supported by an increasing number of experimental
reports \cite{3to11}. Two-gap or two-band supercondutivity was first studied
in the 1950s \cite{12} and has now found renewed relevance in MgB$_2$. In
addition, and contrary to many materials or alloys studied earlier, the two
bands in MgB$_2$ have a roughly equal filling factor, opening the possibility
for interesting new phenomena.

Principly, the two-gap superconductivity can be investigated in
the frame of the charged two-condensate Bose system. This system
is described by a Ginzburg-Landau model with two flavors of Cooper
pairs. Alternatively, it relates to a Gross-Pitaevskii functional
with two charged condensates of tightly bound fermion pairs, or
some other charged bosonic fields. Such theoretical models have a
wide range of applications and have been previously considered in
connection with two-band superconductivity
\cite{11,22,egor2,egor3}. By making use of this theoretical model,
vortices with fractional flux in two-gap superconductors has been
presented \cite{egor2}.

The works mentioned above stimulate us to investigate the topological
structure of the charged two-condensate Bose-Einstein system in more detail.
Let us consider a Bose-Einstein system with two electromagnetically
coupled, oppositely charged condensates, this system can be described
by a two-flavor Ginzburg-Landau-Gross-Pitaevkii (GLGP) functional \cite{egor1},
\begin{eqnarray}
F\!&=&\frac1{2m_1} \left| \left( \hbar \partial _{\mu}
+
i \frac{2e}c A_{\mu} \right ) \Psi_1 \right |^2 \nonumber \\
&+ &\!\frac1{2m_2}\! \left|\!
\left( \hbar \partial _{\mu}\! -\!
i \frac{2e}c A_{\mu} \right)\! \Psi_2\! \right|^2\! +\!
V(\Psi _{1,2})\! +\! \frac{{\bf B}
^2}{8\pi},
\label{glgp}
\end{eqnarray}
in which
\begin{equation}
V(\Psi _{1,2})=-b_{\alpha}|\Psi _{\alpha}|^2 + \frac{c_{\alpha}}2
|\Psi _{\alpha}|^4+ \eta[\Psi _1^* \Psi _2 +\Psi_2^* \Psi_1],
\end{equation}
where $\eta$ is a characteristic of interband Josephson coupling
strength \cite{egor12}. What is much more important in the present
GLGP model is that the two  charged fields are not independent but
nontrivially coupled through the electromagnetic field and the
Josephson coupling. This kind of nontrivial coupling indicates
 that in this system there should be a nontrivial, hidden
topology which, however, cannot be recognized obviously in the
form of Eq. (\ref{glgp}). In order to find out the topological
structure and to investigate it conveniently, we need to reform
the GLGP functional.

By introducing a new set of variables $\rho$ and $\chi _{1,2}$ by
\begin{equation}
\Psi _{\alpha} = \sqrt{2m_{\alpha}} \rho \chi _{\alpha},
\label{chi}
\end{equation}
where the complex $\chi _{\alpha}=|\chi _{\alpha}|e^{i\varphi _{\alpha}}$
satisfying $|\chi _1|^2+ |\chi _2|^2=1$ and the modulus taking the form
\begin{equation}
\rho ^2 = \frac12 \left( \frac{|\Psi_1|^2}{m_1} +\frac{|\Psi_2|^2}{m_2}
\right ),
\label{rho}
\end{equation}
the original GLGP free energy density (\ref{glgp}) can be represented
as
\begin{eqnarray}
F&=&\frac{\hbar^2 \rho^2}4(\partial \vec n) ^2 + \hbar^2 (\partial \rho)^2
+\frac{\hbar^2 c^2}{512 \pi e^2} \Bigg (\frac1{\hbar}[\partial_{\mu}C_{\nu}-
\partial _{\nu} C_{\mu}] \nonumber \\
& &- \vec n \cdot \partial _{\mu} \vec n \times \partial _{\nu} \vec n
\Bigg )^2 + \frac{\rho^2}{16} \vec C ^2 + V(\rho,n_1,n_3),
\label{functional}
\end{eqnarray}
where
\begin{equation}
C_{\mu} =2i\hbar[\chi _1\partial _{\mu} \chi _1^* -\chi _1^*\partial _{\mu} \chi _1
-\chi _2\partial _{\mu} \chi _2^* + \chi _2^*\partial _{\mu} \chi _2]- \frac{8e}c A_{\mu}
\label{cmu}
\end{equation}
\begin{equation}
\vec n = (\bar{\chi}, \vec{\sigma} \chi),
\label{n}
\end{equation}
in which $(\;\; ,\; \;)$ denotes the scalar product and
$\bar{\chi}=(\chi_1^* \;\;\chi _2^*)$, $\vec{\sigma}$ stand for
the Pauli matrices. Now we find that there exists an exact
equivalence between the two-flavor GLGP model and the nonlinear
O(3) $\sigma$ model \cite{fad} which is much more important to
describe the topological structure in high energy physics. By
analogy with it, the counterpart in condensed matter has been
discussed which showed that there are topological excitations in
the form of stable, finite length knotted closed vortices in
two-condensate Bose system \cite{egor1}. In this paper, we will
show that, beside the knotted vortices,
 another kind of topological defects, namely the magnetic
monopoles, also possibly exist in this system. This kind of topological
defects, magnetic monopoles, has also been discussed in chiral supercondoctors
and superfluids \cite{volovik}.

As shown in Eq.(\ref{functional}), we know that the magnetic field
of the system can be divided into two parts, one is the
contribution of field $C_{\mu}$, from Eq.(\ref{cmu}) we learn that
this part is introduced by the supercurrent density \cite{egor1}
and can only present us with the topological defects named
vortices, as what in the single-condensate system. Another part,
the contribution $\vec n \cdot \partial _{\mu} \vec n \times
\partial _{\nu} \vec n$ to the magnetic field term in
Eq.(\ref{functional}), is a fundamentally important property of
the two-condensate system which has no counterpart in a single
condensate system. Indeed, it is exactly due to the presence of
this term that the two-condensate system acquires properties which
are qualitatively very different from those of a single-condensate
system: This term describes the magnetic field that becomes
induced in the system due to a nontrivial electromagnetic
interaction between the two condensates. Thus it is needed to
investigate this induced magnetic field in detail.

The induced magnetic field $\tilde{B}_{\mu}$ is expressed as
\begin{equation}
\tilde{B}_{\mu}= \frac{\hbar c}{8 e}\epsilon _{\mu \nu \lambda}
\epsilon _{abc}n^a \partial_{\nu}n^b \partial _{\lambda} n^c.
\end{equation}
Let us now investigate the divergence of the induced magnetic field
$\tilde{B}_{\mu}$,
namely $Q$, which can be represented in terms of the unit vector field $n^a$ as
\begin{equation}
Q= \partial_{\mu}\tilde{B}_{\mu}=\frac{\hbar c}{8 e}\epsilon _{\mu \nu \lambda}
\epsilon _{abc}\partial_{\mu}n^a \partial_{\nu}n^b \partial _{\lambda} n^c.
\end{equation}
Usually, a unit vector can be written as $\vec n =(\sin \theta
\cos \gamma, \sin \theta \sin \gamma, \cos \theta)$, which was
adopted in many previous works. With this normal expression, one
can immediately draw the conclusion that the divergence of the
induced magnetic field is equal to zero, as the Maxwellian
equation shows.

In fact, in the present context, the unit vector is defined by Eq.
(\ref{n}), from which it is very easy to get
$||\vec{n}||^2=(|\chi_1|^2+|\chi_2|^2)^2$, and the components of
the unit vector $\vec{n}$ are determined by $\chi_{\alpha}$. The
condition $|\chi_1|^2+|\chi_2|^2=1$ we mentioned below Eq.
(\ref{chi}) safeguards that the vector $\vec{n}$ is a unit vector
in the whole space and there is no point at which this condition
can be destroyed. Now let us consider about the components of
field $\chi_{\alpha}$, as well as the components of the unit
vector $n^a$. Normally, the components $\chi_{\alpha}$ are well
defined by Eq. (\ref{chi}), so are the components $n^a$. In this
case, the unit vector field $n^a$ can be reduced to the normal one
and the result of the divergence $Q$ is equal to zero, just as
mentioned above. However, when we investigate the behavior of the
original system described in Eq. (\ref{glgp}) at the points where
both $\Psi_1$ and $\Psi_2$ are zero, the situation is expected to
be changed. After reforming the GLGP functional (\ref{glgp}) via
$\rho$ and $\chi_{1,2}$, we find that even though the condition
$|\chi_1|^2+|\chi_2|^2=1$ is still held, the components
$\chi_{\alpha}$ are not well defined at these points, this is
clear from Eqs.(\ref{chi}) and (\ref{rho}). The same situation
happens to $\vec{n}$, at these points, the unit vector condition
of $\vec{n}$ is still valid, however the components of the unit
vector $n^a$ are undetermined. This indetermination indicates that
these zero points are the topological singular points of the
system, and it leads to an unusual behavior of $\tilde{B}_{\mu}$
and $Q$ at these points. Hence, in order to investigate the
unusual property of $Q$ at these special points, we represent the
unit vector $n^a$ as
\begin{equation}
n^a=\frac{\phi^a}{||\phi||},\;\;\;||\phi||=\sqrt{\phi^a\phi^a}.
\label{phi}
\end{equation}
As discussed above, this is a reasonable representation, $\phi^a$
is a three component vector field related to the order parameters
$\Psi_{\alpha}$ via Eqs.(\ref{n}) and (\ref{chi}). Obviously, it
can be looked upon as a smooth mapping between the
three-dimensional space $X$ (with the local coordinate $x$) and
the three-dimensional Euclidean space $R^3$ $\phi:x \mapsto \vec
\phi (x) \in R^3$, and
 $n^a$  a section of the sphere bundle $S(X)$. Clearly, the zero points
of $\phi^a(x)$ correspond to the points where $\Psi_1$ and
$\Psi_2$ are zero, We would like to name these points as the
singular points of $n^a$. In the following, by virtue of the so
called $\phi$-mapping method \cite{jiang1,jiang2}, we will show
that, when $\phi^a$ field possesses several zero points, the
divergence of the induced magnetic field is no longer zero but
takes the form of the $\delta$-function.

From (\ref{phi}), we have
\begin{equation}
\partial_{\mu} n^a=\frac1{||\phi||}\partial _{\mu}\phi^a + \phi^a
\partial _{\mu}\left(\frac1{||\phi||} \right).
\end{equation}
Due to these expressions, the divergence $Q$ can be rewritten as
\begin{equation}
Q= -\frac{\hbar c}{8 e}\epsilon _{\mu \nu \lambda}
\epsilon _{abc}\partial_{\mu}\phi^d \partial_{\nu}\phi^b \partial _{\lambda}
\phi^c \frac{\partial}{\partial \phi^d }\frac{\partial}{\partial \phi^a }
\left(\frac1{||\phi||} \right)
\end{equation}
If we define the Jacobian $D(\phi/x)$ as
\begin{equation}
\epsilon^{a b c} D(\phi/x)=\epsilon _{\mu \nu \lambda}
\partial_{\mu}\phi^a \partial_{\nu}\phi^b \partial _{\lambda}
\phi^c
\end{equation}
and make use of the Laplacian Green function relation in
$\phi$-space,
\begin{equation}
\frac{\partial}{\partial \phi^d }\frac{\partial}{\partial \phi^a }
\left(\frac1{||\phi||} \right)=-4\pi \delta(\vec \phi),
\end{equation}
we do obtain the $\delta$-function structure of the divergence of
the induced magnetic field as is expected,
\begin{equation}
Q=\partial_{\mu}\tilde{B_{\mu}}= \frac{\hbar\pi c}{2 e}
\delta(\vec{\phi})D(\phi/x), \label{q}
\end{equation}
which shows that the divergence of the induced magnetic field $\tilde{B_{\mu}}$
is not equal to zero. This indicates that in the two-condensate Bose system,
there are point-like topological defects (or namely, magnetic monopoles)
located at the zero point of field $\vec{\phi}$.

Then questions are raised naturally:
what is the topological charges of the magnetic monopoles and how to obtain
 the inner structure of $Q$.

From the above, we see that the zero point of $\vec{\phi}$ plays
an important role. We discuss them more deeply. Suppose the
function $\phi^a(x)$ possesses $l$ isolated zeroes. The implicit
function theorem expresses that when these zeroes are regular
points of $\phi$-mapping and require the Jacobian
$D(\phi/x)\neq0$, the zero points can be expressed by
\begin{equation}
\vec x=\vec z_i,\;\;\;i=1,...,l.
\end{equation}

Now, we will investigate the topological charges of the magnetic monopoles
 and their quantization. Let $M_i$ be a neighborhood of $\vec z_i$ with
boundary $\partial M_i$ satisfying $\vec z_i \not\in \partial M_i$,
$M_i \cap M_j = \emptyset$. Then the generalized winding number $W_i$ of
$n^a(\vec x)$ at $\vec z_i$ can be defined by the Gauss map \cite{jiang3,25}
 $n$: $\partial M_i
\rightarrow S^2$,
\begin{equation}
W_i =\frac1{8\pi}\int _{\partial M_i} n^*(\epsilon _{abc} n^a dn^b
\wedge dn^c ),
\end{equation}
where $n^*$ is the pull back of map $n$. The generalized winding number is
a topological invariant and is also called the degree of Gauss map. It is well
known that $W_i$ are corresponding to the second homotopy group $\pi_2 [S^2]=Z$
(the set of integers). Using the Stokes' theorem in exterior differential form
and the result in (\ref{q}), we get
\begin{equation}
W_i=\int_{M_i} \delta(\vec \phi)D(\phi/x)d^3x.
\label{w}
\end{equation}
By analogy with the procedure of deducing $\delta(f(x))$, one can expand
the $\delta$-function $\delta(\vec \phi)$ as
\begin{equation}
\delta(\vec \phi)=\sum_{i=1}^l c_i \delta(\vec x - \vec z_i),
\label{delta}
\end{equation}
where the coefficients $c_i$ must be positive, i.e. $c_i=|c_i|$. Substituting
(\ref{delta}) into (\ref{w}) and calculating the integral, we get an
expression for $c_i$,
\begin{equation}
c_i = \frac{|W_i|}{|D(\phi/x)_{\vec x = \vec z_i}|}.
\end{equation}
Letting $|W_i| = \beta _i$, using this expansion of $\delta (\vec
\phi)$, it is evident that $Q$ in (\ref{q}) can be further
expressed in the form
\begin{equation}
Q = \frac{\hbar\pi c}{2 e} \sum _{i=1}^l \beta _i \eta _i \delta (\vec x
-\vec z_i),
\end{equation}
where the positive integer $\beta _i$ is the so-called Hopf index of the
$\phi$-mapping on $M_i$, $\eta _i= {\rm sign}D(\phi/x)_{\vec z_i} = \pm1$
is the Brouwer degrees of the $\phi$-mapping. This expression
 is exactly the density
of the system of $l$ topological defects (magnetic monopoles) with
corresponding topological charges
$g_i =\beta_i \eta _i$, the Hopf indices $\beta _i$
characterize the absolute value of the monopole topological charges and the
Brouwer degrees $\eta_i =1$ correspond to monopoles while $\eta _i=-1$ to
antimonopoles.

In summary, by making use of the $\phi$-mapping method, we find that,
beside the closed knotted vortices \cite{egor1}, there is another
kind of topological defects, namely the magnetic monopoles, in two-band
superconductors. Moreover, these monopoles are located at the zero points
of the field $\vec{\phi}$, or namely, at the singular points of the field
$\vec{n}$ defined in Eq.(\ref{n}), and their topological charges can be
expressed in terms of the Hopf indices and the Brouwer degrees of the
$\phi$--mapping.  This is very similar with a different physical system
discussed by Cho \cite{cho1}.

\vskip0.3cm

The author (Y.J.) gratefully acknowledges Professor R. Ikeda for
bringing some relevant references to his attention.
 This work was supported in part by
the Alexander von Humboldt Foundation.

\end{document}